\documentclass[sigconf,table,xcdraw,screen]{acmart}
\usepackage{booktabs}
\usepackage{amsmath}
\usepackage{tikz}
\usepackage{pgfplots}
\usepackage{amssymb}
\usepackage{amsfonts}
\usepackage{algorithmic}
\usepackage{graphicx}
\usepackage{listings}
\usepackage{textcomp}
\usepackage{multirow}
\usepackage{xspace}
\usepackage{paralist}
\usepackage{enumitem}
\usepackage{comment}
\usepackage{flushend}
\usepackage{tikz}
\usepackage{todonotes}
\usepackage{graphicx}
\usepackage{caption}
\usepackage{subcaption}
\usepackage{amssymb}
\usepackage{ifthen}
\usepackage{wrapfig}
\usepackage{xcolor}
\usepackage{listings}
\usepackage{url}
\usepackage{hyperref}
\usepackage{titlesec}
\usepackage{multirow}
\usetikzlibrary{arrows,calc,shapes}

\newcommand{\ie}{\textit{i.e.,}\xspace}
\newcommand{\eg}{\textit{e.g.,}\xspace}
\newcommand{\etc}{\textit{etc.}\xspace}
\newcommand{\etal}{\textit{et al.}\xspace}

\newcommand{\secref}[1]{Section~\ref{#1}\xspace}

\newcommand{\figref}[1]{Figure~\ref{#1}\xspace}

\newcommand{\tabref}[1]{Table~\ref{#1}\xspace}

\newcommand{\atlas}{{\sc Atlas}\xspace}

\AtBeginDocument{%
  \providecommand\BibTeX{{%
    \normalfont B\kern-0.5em{\scshape i\kern-0.25em b}\kern-0.8em\TeX}}}

\copyrightyear{2020}
\acmYear{2020}
\setcopyright{acmcopyright}
\acmConference[ICSE '20]{42nd International Conference on Software
Engineering}{May 23--29, 2020}{Seoul, Republic of Korea}
\acmBooktitle{42nd International Conference on Software Engineering (ICSE '20), May
23--29, 2020, Seoul, Republic of Korea}
\acmPrice{15.00}
\acmDOI{10.1145/3377811.3380429}
\acmISBN{978-1-4503-7121-6/20/05}

\begin{document}

\title{On Learning Meaningful Assert Statements for Unit Test Cases}



\author{Cody Watson}
\affiliation{%
  \institution{Washington and Lee University}
  \city{Lexington}
  \state{Virginia}
  \postcode{24450}
}
\email{cwatson@wlu.edu}

\author{Michele Tufano}
\affiliation{%
  \institution{Microsoft}
  \city{Redmond}
  \state{Washington}
  \postcode{98052}
}
\email{michele.tufano@microsoft.com}

\author{Kevin Moran}
\affiliation{%
  \institution{William \& Mary}
  \city{Williamsburg}
  \state{Virginia}
  \postcode{23185}
}
\email{kpmoran@cs.wm.edu}

\author{Gabriele Bavota}
\affiliation{%
  \institution{Università della Svizzera italiana (USI)}
  \city{Lugano}
  \state{Switzerland}
}
\email{gabriele.bavota@usi.ch}

\author{Denys Poshyvanyk}
\affiliation{%
  \institution{William \& Mary}
  \city{Williamsburg}
  \state{Virginia}
  \postcode{23185}
}
\email{denys@cs.wm.edu}

\begin{abstract}
Software testing is an essential part of the software lifecycle and requires a substantial amount of time and effort. It has been estimated that software developers spend close to 50\% of their time on testing the code they write. For these reasons, a long standing goal within the research community is to (partially) automate software testing. While several techniques and tools have been proposed to automatically generate test methods, recent work has criticized the quality and usefulness of the assert statements they generate. Therefore, we employ a Neural Machine Translation (NMT) based approach called \atlas (\underline{A}u\underline{T}omatic \underline{L}earning of \underline{A}ssert \underline{S}tatements) to automatically generate meaningful assert statements for test methods. Given a test method and a focal method (\ie the main method under test), \atlas can predict a meaningful assert statement to assess the correctness of the focal method. We applied \atlas to thousands of test methods from GitHub projects and it was able to predict the exact assert statement manually written by developers in 31\% of the cases when only considering the top-1 predicted assert. When considering the top-5 predicted assert statements, \atlas is able to predict exact matches in 50\% of the cases. These promising results hint to the potential usefulness of our approach as (i) a complement to automatic test case generation techniques, and (ii) a code completion support for developers, who can benefit from the recommended assert statements while writing test code.  %Moreover, we show that generated assert statements that are not identical to what the developer wrote, still provide meaningful insight into how to effectively test the focal method. %Given a test method and the focal method it tests in the production code, we learn the appropriate context to generate a meaningful assert statement. We show that ATLAS can recommend complex and meaningful (perfect) assert statements in 31\% of cases, with no added burden to the developer. Additionally, we can accurately predict up to 50\% when the developer considers the top-5 generated assert statements, which only requires minimal effort from the developer. 
%\DENYS{I would add here another sentence saying: "Another X\% of automatically generated asserts needed only minor modifications." This would be based on the analysis results of less than perfect asserts.} \MICHELE{I agree with Denys, we could nicely complete this sentence adding more. But rather than mentioning the the "less than perfect asserts", I would mention here the incredible beam=5 results! "While in 50\% of the cases the correct assert statement is in the top-5 suggestions". Basically here the only extra work for the developer is to pick the right one from a small list of 5 suggestions.} \DENYS{The two sentences before with 31\% and 50\% need more work. I feel like the message is not as effective as it can be. I will keep thinking.}
%Moreover, we used our approach to generate assert statements for test methods within open pull requests on GitHub, demonstrating the use and practicality of our approach in a real-world software development scenario.
\vspace{-0.3cm}
\end{abstract}

\begin{CCSXML}
<ccs2012>
<concept>
<concept_id>10010520.10010553.10010562</concept_id>
<concept_desc>Computer systems organization~Embedded systems</concept_desc>
<concept_significance>500</concept_significance>
</concept>
<concept>
<concept_id>10010520.10010575.10010755</concept_id>
<concept_desc>Computer systems organization~Redundancy</concept_desc>
<concept_significance>300</concept_significance>
</concept>
<concept>
<concept_id>10010520.10010553.10010554</concept_id>
<concept_desc>Computer systems organization~Robotics</concept_desc>
<concept_significance>100</concept_significance>
</concept>
<concept>
<concept_id>10003033.10003083.10003095</concept_id>
<concept_desc>Networks~Network reliability</concept_desc>
<concept_significance>100</concept_significance>
</concept>
</ccs2012>
\end{CCSXML}

\ccsdesc[300]{Software Testing~Unit Tests}
\ccsdesc[300]{Artificial Intelligence~Machine Translation}
\vspace{-0.1cm}
%\keywords{Testing, Unit Tests, Deep Learning, Neural Machine Translation}

\maketitle

\section{Introduction}
\label{sec:intro}
% Introduction Page

Writing high-quality software tests is a difficult and time-consuming task. To help tame the complexity of testing, ideally, development teams should follow the prescriptions of the test automation pyramid \cite{Cohn:2009}, which suggests first writing \textit{unit tests} that evaluate small, functionally discrete portions of code to spot specific implementation issues and quickly identify regressions during software evolution. Despite their usefulness, prior work has illustrated that once a project reaches a certain complexity, incorporating unit tests requires a substantial effort in traceability, decreasing the likelihood of unit test additions~\cite{7107469}. Further challenges exist for updating existing unit tests during software evolution and maintenance~\cite{7107469}.

To help address these issues the software testing research community has responded with a wealth of research that aims to help developers by automatically generating tests~\cite{ESECFSE11,radoop}. However, recent work has pointed to several limitations of these automation tools and questioned their ability to adequately meet the software testing needs of industrial developers~\cite{7965450,Shamshiri:FSE'15}. For example, it has been found that the \textit{assert statements} generated by state-of-the-art approaches are often incomplete or lacking the necessary complexity to capture a designated fault. \textbf{The generation of meaningful assert statements is one of the key challenges in automatic test case generation}. Assert statements provide crucial logic checks in a test case to ensure that the program is functioning properly and producing expected results. However, writing or generating effective assert statements is a complex problem that requires knowledge pertaining to the purpose of a particular unit test and the functionality of the related production code. Thus, an effective technique for the generation of assert statements requires predicting both the type and logical nature of the required check, using source and test code as contextual clues for prediction.

To help advance techniques that aid developers in writing or generating unit tests, we designed \atlas, an approach for automatically generating syntactically and semantically correct unit test assert statements using Neural Machine Translation (NMT). \atlas generates models trained on large-scale datasets of source code to accurately predict assert statements within test methods. We take advantage of the deep learning strategy of NMT, which has become an important tool for supporting software-related tasks such as bug-fixing \cite{Tufano:2018:EIL:3238147.3240732,Chen:2019,Mesbah:2019:DLR:3338906.3340455,DBLP:journals/corr/abs-1812-07170}, code changes \cite{DBLP:journals/corr/abs-1901-09102}, code migration \cite{Nguyen:2014:MCS:2591062.2591072,Nguyen:2013:LSM:2491411.2494584}, code summarization \cite{LeClair:2019:NMG:3339505.3339605,Liu:2018:NCM:3238147.3238190,inproceedingsMine}, pseudo-code generation \cite{Oda:2015:LGP:3343886.3343959}, code deobfuscation \cite{Vasilescu:2017:RCN:3106237.3106289,Jaffe:2018:MVN:3196321.3196330} and mutation analysis \cite{DBLP:journals/corr/abs-1812-10772}. To the best of our knowledge, this is the first empirical step toward evaluating an NMT-based approach for the automatic generation of assert statements. Specifically, we embed a test method along with the context of its focal method \cite{5609581} (\ie a declared method, within the production code, whose functionality is tested by a particular assert statement) and we "translate" this input into an appropriate assert statement. Since our model only requires the test method and the focal method, we are able to aid developers in automatic assert generation even if the project suffers from a lack of initial testing infrastructure. Note that our approach is not an alternative to automatic test case generation techniques~\cite{ESECFSE11,radoop}, but rather, a complementary technique that can be combined with them to improve their effectiveness. In other words, the automatic test case generation tools can be used to create the test method and our approach can help in defining a meaningful assert statement for it.

To train \atlas, we mined GitHub for every Java project making use of the JUnit assert class. In total we analyzed over 9k projects to extract 2,502,623 examples of developer-written assert statements being used within test methods. This data was used to give the NMT model the ability to generate assert statements that closely resemble those created by developers. Therefore, not only do we enable efficiency within the software testing phase but we also facilitate accuracy and naturalness by learning from manually written assert statements. After we extracted the pertinent test methods containing assert statements from the Java projects, we automatically identified the focal method for each assert statement based on the intuition from Qusef \etal \cite{5609581}. We hypothesize that combining the test method and the focal method should provide the model with enough context to automatically generate meaningful asserts.

We then \textit{quantitatively} and \textit{qualitatively} evaluated our NMT model to validate its usefulness for developers. For our \textit{quantitative} analysis, we compared the models generated assert statements with the oracle assert statements manually written by developers. We considered the model successful if it was able to predict an assert statement which is identical to the developer-written one. Our results indicate that \atlas is able to automatically generate asserts that are identical to the ones manually written by developers in 31.42\% of cases (4,968 perfectly predicted assert statements) when only considering the top-1 predicted assert. When looking at the top-5 recommendations, this percentage rises to 49.69\% (7,857).

For our \textit{qualitative} analysis we analyzed ``imperfect predictions'' (\ie predictions which can differ semantically or syntactically as compared to the assert manually written by developers) to understand whether they could be considered an acceptable alternative to the original assert. We found this to be true in the 10\% of cases we analyzed. Finally, we computed the edit distance between the imperfect predictions and original asserts in order to assess the effort required for a developer to adapt a recommended assert statement into one she would use. We show that slight changes to the ``imperfect'' asserts can easily convert them into a useful recommendation. % for developers. %For our qualitative analysis, we generate assert statements for test methods within active pull requests on GitHub. We then await feedback from the developers either in the form of a merge or in the form of comments about the quality of the suggested assert statements. 
To summarize, this paper provides the following contributions:
\vspace{-0.1cm}
\begin{itemize}
	\item We introduce \atlas, a NMT-based approach for automatically generating assert statements. We provide details pertaining to the mining, synthesizing, and pre-processing techniques to extract test methods from the wild, and to train and test \atlas; 
	\item An empirical analysis of \atlas and its ability to use NMT to accurately generate a semantically and syntactically correct assert statement for a given test method;
	\item A quantitative evaluation of the model, and a detailed comparison between modeling raw and abstracted test methods;
	\item A publicly available replication package \cite{replication} containing the source code, model, tools and datasets discussed in this paper.
\end{itemize}

%\MICHELE{@Cody and @All I have a suggestion, but feel free to keep as it is. I believe we should avoid giving too much credits to NMT (\eg "ability of NMT to ..."). I would actually suggest to try to avoid mentioning always NMT, but rather mentioning our approach. Our approach is not "just" NMT. We intelligently organize the test method and focal method, we perform code abstraction, we choose particular token as placeholder for the assert, and then we generate it using Encoder-Decoder model. What I described is not NMT, it's our approach :) @Cody, do you want to give a name to it?} \CODY{Yes, I agree with Michele. I changed the introduction to reflect this, I like the name but I'm open to suggestions}

\section{Related Work \& Motivation}
\label{sec:motivation}
% Realted Work & Motivation Page

%\GABRIELE{Commented out the following since it is a repetition of the introduction}
%Although unit tests are a fundamental step in the process of software testing, the ability to test individual components of functionality is a crucial step toward more complex types of software testing. Assert statements provide the ability to capture the logic and functionality of the underlying code base within unit tests. Therefore, generating meaningful assert statements is necessary to detect potential faults. Despite the importance of unit tests, they can be incredibly time consuming and a laborious process for the developer. In order to reduce this burden on developers, automated tools such as Evosuite \cite{Fraser:ESEC/FSE'11}, Randoop \cite{radoop} and Agitar \cite{agitar} will automatically generate entire test methods.

While several automated tools such as Evosuite \cite{Fraser:ESEC/FSE'11}, Randoop \cite{radoop} and Agitar \cite{agitar} have been proposed to automatically generate test methods, these tools embed their own methods for synthesizing assert statements. Evosuite, one of the most popular automated test generation tools, uses a mutation-based system to generate appropriate assert statements. In particular, it introduces mutants into the software and attempts to generate assert statements able to kill these mutants. Evosuite also tries to minimize the number of asserts while still maximizing the the number of killed mutants~\cite{Fraser:ESEC/FSE'11}. 

Randoop is another automated test generation tool that creates assertions with \textit{intelligent} guessing. This technique applies feedback-directed random testing by analyzing execution traces of the statement it creates. Essentially, a list of \textit{contracts}, or pieces of logic that the code must follow, are used to guide the generation of assert statements. These contracts are very similar to user defined assert statements. However, the contracts only provide the logic. The Randoop program creates a syntactically correct assert statement that tests the user's provided logic pertaining to the test method. Differently from Evosuite and Randoop, \atlas applies a deep learning-based approach, with the goal of mimicking the behavior of expert developers when writing assert statements.

Recent research has evaluated the ability of state-of-the-art automated techniques for test generation to capture real faults~\cite{Shamshiri:FSE'15,7965450}. %and found that the quality of the generated assert statements are one of primary reasons that automatically generated tests failed to uncover real faults. 
The recent study conducted by Shamshiri \etal~\cite{Shamshiri:FSE'15} highlights the importance of high-quality, complex assert statements when detecting real faults. The authors compare the abilities of both exceptions and asserts as fault finding mechanisms. They note that three automated approaches (Evosuite, Randoop, Agitar) detect more faults through the use of an assert statement, however, insufficiencies in the automatically generated asserts led to many bugs going undetected~\cite{Shamshiri:FSE'15}. Thus, this work makes two important conclusions: (i) assert statements are an important component of automated test case generation techniques, as they are the main vehicle through which faults are detected; (ii) the current quality of assert statements generated by automated testing techniques are often not of high enough quality to detect real faults. %This work concludes that the generated asserts are either incomplete or lacking the necessary complexity to capture the designated fault, but assert statements are indeed important, as they are the main modality through which faults are detected.

A complimentary study, performed by Almasi \etal \cite{7965450}, tested Evosuite and Randoop's ability to detect 25 real faults in an industrial software system. %developed by SEB Life \& Pension Holding AB Riga Branch. 
The study found that Evosuite was able to detect 56.4\% and Randoop was able to detect 38.0\% of the faults within that system. Of particular note was the author's qualitative evaluation, which showed that nearly half of the undetected faults could have been detected with a more appropriate assert statement. Additionally, the authors solicited feedback from developers, asking "\textit{How can the generated tests be improved?}". The general consensus among the respondents was that automated testing strategies failed to generate meaningful assert statements. The authors also presented hand-written test methods to developers for comparison, and they commented that "\textit{the assertions are meaningful and useful \textbf{unlike the generated ones}}". These findings demonstrate a clear need for automated techniques that generate \textit{meaningful} assert statements to complement existing test generation approaches. %\MICHELE{Very nice motivations and quotes. If we have space, later on we could think to highlight/boxed/quote some parts of this paragraph.}

One of the limitations that tools such as Evosuite, Randoop and Agitar have is that they rely on heuristics or "intelligent" randomness in order to generate assert statements. This type of generation does not take into account the learn-able components of test and focal methods, and thus leads to more simplistic asserts. Therefore, we leverage a NMT-based DL technique to generate asserts that can test the complexities contained within the context of the test and focal method. This strategy results in an assert  which possesses the ability to accurately evaluate the logic of the focal method, leading to more useful unit tests and a higher quality test suite.

\vspace{-0.2cm}
\section{A{\normalsize TLAS}: Learning Asserts via NMT}
\label{sec:approach}
% Approach Page

\begin{figure}
	\centering
	\vspace{0.2cm}
	\includegraphics[width=0.82\columnwidth]{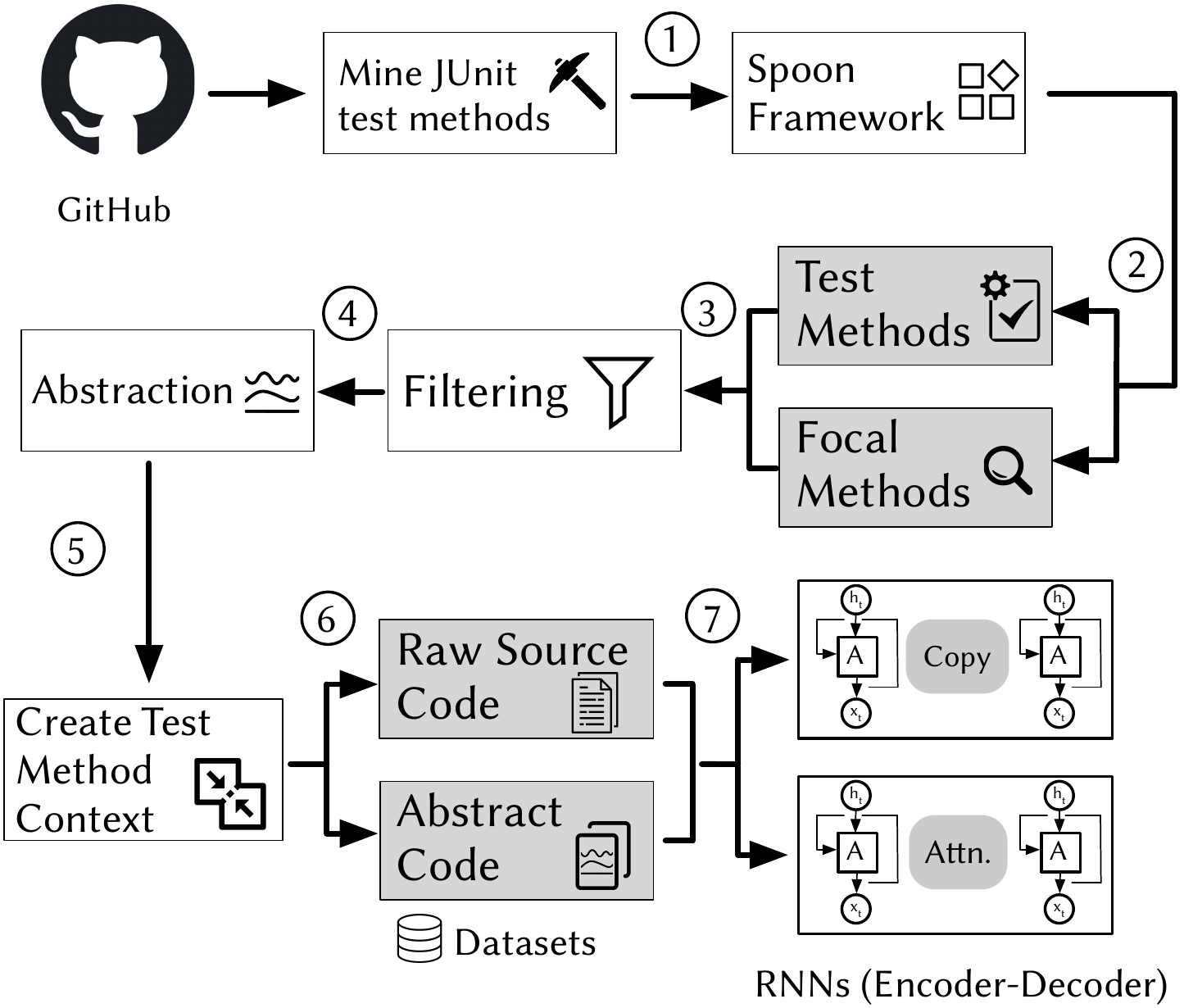}
	\vspace{-0.2cm}
	\caption{Overview of A{\footnotesize TLAS} Workflow} %\MICHELE{I was sketching an image for the approach, but then realized that Cody already created it, but not inserted in the paper. I enabled it. I think it's good! I believe this is way more informative than the Encoder-Decoder one. I would actually change the last section, showing two Encoder-Decoder models, the one with Copy Mechanism for the raw source code, and the other one without copy mechanism for the abstracted code. We can drop details about the RNN and just keep high-level design. MINOR: I'm not a big fan of this blue color, but that's just up to you Cody :)}
	\label{fig:approach}
\vspace{-0.5cm}
\end{figure}

We provide an overview of the \atlas workflow for learning assert statements via NMT in Fig.~\ref{fig:approach}. Our approach begins with the mining and extraction of test methods from Java projects. To do this, we mine GitHub projects that use the JUnit testing framework (Sec. \ref{sec:approach:github}). From those projects, we mine all test methods denoted with the \texttt{@Test} annotation as well as every declared method within the project (Sec. \ref{sec:approach:extraction}). We then filter this data, identify the appropriate focal method context, and generate pairs containing the contextual test method (\ie the test method augmented with information about the focal method it tests) and the relevant assert statement (Sec. \ref{sec:approach:focal} \& Sec. \ref{sec:approach:filtering}). We refer to these pairs as Test-Assert Pairs (TAPs). Next, we generate two datasets of TAPs: (i) Raw Source Code, where TAPs are simply tokenized; (ii) Abstract Code, where we abstract the TAPs through our abstraction process (Sec. \ref{sec:approach:taps}). Finally, we train two RNN encoder-decoder models; one using the copy mechanism trained on the Raw Source Code TAPs, and another using only the attention mechanism trained on the Abstract Code TAPs (Sec. \ref{sec:approach:seq2seq}).

%Finally, we abstract the TAPs through our abstraction process and organize the data into training, validation and testing sets (Sec. \ref{sec:approach:taps}).

\subsection{GitHub Mining}
\label{sec:approach:github}
Our main motivation toward studying Java projects that use the JUnit framework is \textit{applicability}. As of August 2019 the TIOBE Programming Community Index indicated Java as the most popular programming language \cite{tiobe}. In addition, a study done by Oracle in 2018, found that JUnit was the most popular Java library \cite{poirier}. Hence, curating a dataset of projects that use Java and JUnit lends to the potential for impact on real-world software development.

We identified GitHub projects using the JUnit testing framework. Since client projects can use JUnit by declaring a dependency through Apache Maven \cite{maven}, we started by using the GitHub search API to identify all Java projects in GitHub having at least one \texttt{pom} file needed to declare dependencies toward Maven libraries. This resulted in the identification of 17,659 client projects, using 118,626 \texttt{pom} files and declaring $\sim$1.1M dependencies in total. We downloaded all the identified \texttt{pom} files and mined them to identify all client projects declaring a dependency toward JUnit version 4 and all its minor releases. These dependencies can be easily identified by looking in the \texttt{pom} file for artifacts having the \texttt{junit} \emph{groupId}, \texttt{junit} \emph{artifactId}, and a version starting with ``4.''. Using this process, we collected a total of 9,275 projects. Note that we decided to focus on JUnit v.4 since, in the mined dataset of \texttt{pom} files, we found that the majority of them had a dependency towards this version. 

\vspace{-0.2cm}
\subsection{Method Extraction}
\label{sec:approach:extraction}
After mining these projects from GitHub, we downloaded the source code and extracted the relevant test methods using Spoon \cite{Pawlak:2016:SLI:3035074.3035076}. This framework allows for source code analysis through the creation of a meta-model where the user can access program elements. To access relevant methods, we extract methods beginning with the \texttt{@Test} annotation, which is inherent to the JUnit framework. After extracting the test methods, we extract every method declared within the project, excluding methods from third party libraries. The extracted methods comprise a pool, from which we can determine the focal method of interest for a particular test method. The reason we only consider methods declared within the project is two-fold. First, most assert statements are evaluating the internal information of the project itself rather than information taken from third party libraries or external packages. Second, it would require a substantial effort to retrieve the method bodies and signatures from all the third party libraries and external packages. Since our goal is to learn appropriate assert statements for a given test method and its context, any test method without an assert statement has been discarded. Also, since this is the first work in the literature applying NMT to automatically generate assert statements, we decided to focus on test methods having a single assert statement and, thus, we exclude those implementing multiple asserts. While we acknowledge that this is a simplification of the problem we tackle, we preferred to first investigate the ``potential'' usefulness of NMT in a well-defined scenario in which, for example, it is safe to assume that the whole test method provides contextual information for the unique assert statement it contains. This assumption is not valid in the case of multiple asserts, and instead requires the development of techniques which are able to link parts of the test method body to the different asserts in order to understand the relevant context for each of them. This is part of our future work. Overall, we collected 188,154 test methods with a single assert statement.

\subsection{Identifying Focal Methods}
\label{sec:approach:focal}
Our next task is to identify the focal method that the assert statement, within the test method, is testing. To accomplish this we implement a heuristic inspired by Qusef \etal \cite{5609581}. We begin by extracting every method called within the test method. The list of invoked methods is then queried against the previously extracted list of methods defined inside the project, considering the complete method signature. % (\ie path, name, and parameters).
We then assume that the last method call before the assert is the focal method of the assert statement \cite{5609581}. In some instances, the assert statement contains the method call within its parameters. In these cases, we consider the method call within the assertion parameters as the focal method. It may appear problematic that we use the line we attempt to predict in order to extract the focal method (since, in theory, the line to generate should not exist). However, in a real usage scenario, we assume that the developer can provide the focal method to our model (\ie she knows the method she wants to test). Since identifying the focal method manually for a large number of assert statements is unreasonable, we used the heuristic previously described in place of manual identification of the focal method. We note this as a limitation but find it reasonable that either a developer or an automated test generation strategy would provide this information to our approach. %We reiterate that in a real-world development scenario, we rely on the knowledge of the developer to provide the test and focal method of interest. However, for training, we developed a heuristic which uses the assert statement line solely to extract the focal method context.

\subsection{Filtering}
\label{sec:approach:filtering}
In this work, we are attempting to generate semantically and syntactically correct assert statements from the test method and focal method context. Thus, we are creating a model which must learn relationships from source code. Modeling this type of data presents certain challenges, such as the open vocabulary problem and the ``length of the input'' problem \cite{DBLP:journals/corr/KoehnK17}. Usually, these problems are tackled by limiting the vocabulary and the input length so that the model can adequately learn  \cite{DBLP:journals/corr/SutskeverVL14}. We employ similar solutions when training our model. We filter the data in three distinct ways: i) excluding test methods longer than 1,000 tokens; ii) filtering test methods that contain an assert statement which requires the synthesis of unknown tokens; and iii) removing duplicate examples within the dataset. Every filtering step helps to address NMT-related challenges and have been used in previous approaches that take advantage of this deep learning based strategy \cite{DBLP:journals/corr/KoehnK17,Chen:2019,Tufano:2018:EIL:3238147.3240732}. %The first filtering step is fairly straight forward and has been used to 
%\GAB{we need a justification here}. \MICHELE{We can cite this: "Six Challenges for Neural Machine Translation" \cite{DBLP:journals/corr/KoehnK17}, one of it is dealing with long sentences.}

Our first filtering step is fairly straightforward: we remove all test methods that exceed 1,000 tokens. The second filtering step removes test methods in which the appropriate assert statement requires the synthesis of one or more unknown tokens. This means that the syntactically and semantically correct assert statement requires a token that cannot be found in the vocabulary or in the contextual method (\ie test method + focal method). %\GAB{what's the difference between vocabulary and context? Isn't enough to say vocabulary?} \MICHELE{No! They are very different. We should probably give a better intuition here so we avoid reviewers' complaints. The vocabulary is formed by all the tokens that we consider "idioms", so the ones that we do not abstract (top-1k). Tokens in the context (or better: in the Contextual test method, which means Test Method + Tested Method) are all the tokens that appear in those two methods, even if they are not part of our vocabulary. So we can generate asserts that contain tokens that are either in the vocabulary or in the \textbf{current} contextual method. We could also formally define this: Given a test method $tm$ and the corresponding focal method $fm$, we remove TAPs for which the assert statement to generate contains a token $t$ such that $t \notin \{\ V_{global} \cup V_{tm} \cup V_{fm}\}$, where $V$ represent the vocabulary... But probably this type of explanation should happen after the Vocabulary subsection.} 
Indeed, there is no way to synthesize these tokens when the model attempts to generate a prediction. We further explain this problem as well as our developed solution in \secref{sec:approach:vocab}. Lastly, our third filtering step aims at removing duplicated instances, ensuring that every contextual method and assert statement pair in our dataset is unique.

\subsubsection{Vocabulary}
\label{sec:approach:vocab}

%\begin{figure}
%	\centering
%	\includegraphics[width=\linewidth]{fig/vocab_overview.png}
%	\caption{Zipf's Distribution of Vocabulary}
%	\label{fig:vocabulary}
%\end{figure}  

%The vocabulary to a NMT approach is an important pre-condition of the model. 
We have alluded to the open vocabulary problem which is an inherent limitation of NMT. This issue arises because developers are not limited in the number of unique tokens they can use. They are not limited to, for example, English vocabulary, but also create ``new'' tokens by combining existing words (\eg by using the CamelCase notation) or inventing new words to comprise identifiers (\eg V0\_new). For this reason, the source code vocabulary frequently needs to be artificially truncated. To deal with this problem, we studied the tokens distribution in our dataset, observing that it follows Zipf's law, as also found by previous work analyzing open source software lexicons \cite{5090047}. This means that the dataset's tokens follow a power-law like distribution with a long tail, and that many of the TAPs in our dataset can be successfully represented by only considering a small subset of its 695,433 unique tokens. Based on analyzing the data and on previous findings in the literature \cite{DBLP:journals/corr/abs-1901-01808}, we decided to limit our vocabulary to the 1,000 most frequent tokens. This allows us to successfully represent all tokens for 41.5\% of the TAPs in our dataset (\ie 204,317 out of 491,649). %\CODY{Do we need to mention these numbers here? Its a fairly limiting step and I haven't seen other related work mention the specific number of examples they remove doing this technique.} \DENYS{Let's keep the numbers for now and decide later; another alternative is to mention these numbers later in the paper}

This filtering step aimed at removing instances for which the model would need to generate an unknown token. This can be formally defined as follows:  Given a contextual method (\ie the test method $tm$ and the corresponding focal method $fm$), we remove TAPs for which the anticipated assert to generate contains a token $t$ such that $t \notin \{\ V_{global} \cup V_{tm} \cup V_{fm}\}$, where $V$ represents the vocabulary. We refer to $V_{tm} \cup V_{fm}$ as the contextual method. Each filtering step was due to concrete limitations of state-of-the-art NMT models. In numbers, starting from $\sim{750}$ thousand test methods having a single assert, $\sim{2.5}$ thousand tests are removed due to their excessive length, and $\sim{280}$ thousand due to unknown tokens. Thus, $\sim{37\%}$ of the single assert test methods are removed.

\vspace{-0.2cm}
\subsection{Test-Assert Pairs and Abstraction}
\label{sec:approach:taps}
The next step of our approach consists of preparing the data in such a manner that it can be provided as input to \atlas's NMT model. This process involves: i) the concatenation of the focal method to the test method in order to create the Test-Assert Pair (TAP); ii) the tokenization of the TAP; and iii) the abstraction of the TAP. Starting from the first point, it is important to note that not every test method will inherently contain a focal method being tested. However, for the test methods that do posses a focal method, we append its signature and body to the end of the test method. While comments and whitespaces are ignored, we do not remove any other token from either the test or focal method. We then proceed to remove the entire assert statement from the test method, replacing it with the unique token "\texttt{AssertPlaceHolder}". Therefore, the first part of a TAP consists of the concatenated test and focal method, and the second part consists of the assert statement to generate.

We generate two separate datasets of TAPs. The first uses the raw source code to represent the test and the focal method.  The second dataset consists of abstracted TAPs, in which the source code tokens are abstracted to limit the vocabulary and to increase the possibility of observing recurring patterns in the data. As previous studies have shown \cite{Tufano:2018:EIL:3238147.3240732}, this can lead to an increase in performance without losing the ability to automatically map back the abstracted tokens to the raw source code.  

Our abstraction process tokenizes the input, determines the type for each individual token, replaces the raw token with its abstraction and finally creates a map from the abstracted tokens back to the raw ones. Each TAP is abstracted in isolation and has no affect on the way other TAPs are abstracted. We start by using the \texttt{javalang} tool to tokenize the input, which allows us to analyze the type of token. This tool transforms the Java method into a stream of tokens, which is then parsed by the same tool to determine their type (\eg whether a token represents an identifier, a method declaration, \etc) \cite{thunes_2019}. We use these types in order to create an expressive representation that maintains the structure of the code. In total, there are 13 different types of tokens that we use as part of our abstraction process (complete list in our replication package \cite{replication}). When we encounter one of these types, we replace the raw source code token with an abstraction term that indicates the type and number of occurrences of that type of token up to that moment. In other words, we use a numerical value to differentiate tokens of the same type. For example, suppose we encounter a token that is determined to be a method call. This token is replaced with the term METHOD\_0, since this is the first type of that token we have encountered. The next time we encounter a new method call in the same stream of tokens, it would be assigned the term METHOD\_1. In the event where the same token appears multiple times within the TAP, it is given the same abstraction term and numerical value. This means that if the same method is invoked twice in the test or in the focal method and it is represented with the term METHOD\_0, the abstraction will contain ``METHOD\_0'' twice. Since all TAPs are abstracted in isolation, we can reuse these terms when abstracting a new TAP, thus limiting the vocabulary size for our NMT model. 

\begin{figure}
	\centering
	\includegraphics[width=1.04\linewidth]{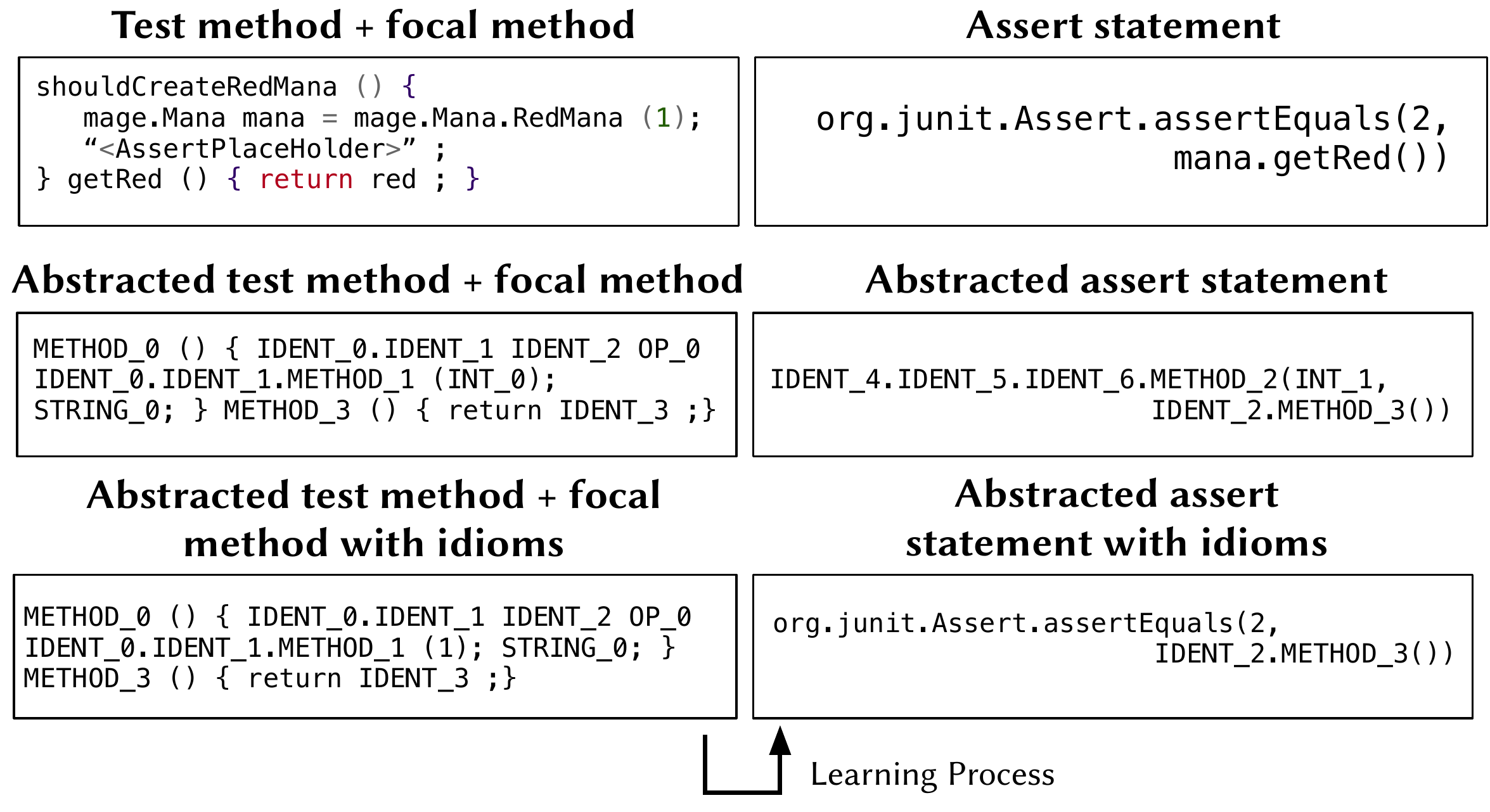}
	\vspace{-0.7cm}
	\caption{Overview of the abstraction process}
	\label{fig:abstraction}
    \vspace{-0.5cm}
\end{figure}      
%\vspace{-0.9cm}

In addition to the abstraction terms that replace the raw source code tokens, we take advantage of idioms in order to create an abstraction that captures more semantic information. The inclusion of idioms can also help contextualize surrounding tokens, since some abstracted tokens may be more likely to appear around a particular idiom. In addition, idioms help to prevent the exclusion of a TAP due to the synthesis of an unknown token. For instance, consider the example of abstraction shown in \figref{fig:abstraction}. In the middle of the figure it is possible to see that INT\_1, IDENT\_4, IDENT\_5 and IDENT\_6 only appear in the abstracted assert statement, but do not appear in the abstracted test and focal method. If we only relied on the abstraction, we would be unable to resolve these tokens that are unique to the predicted assert statement. Therefore, we keep the raw value of common idioms (\ie the top 1,000 tokens in our dataset in terms of frequency) in our abstracted representation, as shown in the bottom part of \figref{fig:abstraction}. 

Overall, the vocabulary of the Abstract TAPs comprises 1,000 idioms plus $\sim$100 typified IDs, while the vocabulary of the Raw Source Code TAPs contains 1,000 tokens.

%The use of idioms slightly increases our vocabulary but allows ATLAS to learn from examples such as the one shown in \figref{fig:abstraction}. %As additional analysis, we empirically compare the performance of ATLAS when using the raw dataset without our abstraction process and the dataset that is subjected to our abstraction process.

%Once the TAPs are created, we then tokenize both the test and focal methods as well as the assert statement. This tokenization is performed using the javalang tool \cite{thunes_2019} and conformed to the format that is accepted by the NMT model where tokens are separated by a single space. 
%\MICHELE{I believe we don't do another tokenization step. We tokenize before the abstraction, and the abstraction process does not mess up the initial tokenization. I would suggest to remove this so we don't confuse the reader.} %We empirically compare performance of NMT when using the raw dataset and the abstracted dataset.

\vspace{-0.2cm}
\subsection{Sequence to Sequence Learning}
\label{sec:approach:seq2seq}
%\MICHELE{We should mention here that this approach is similar to previous ones (cite) and in particular (concat, placeholder, single-line output) is inspired by the latest Monperrous paper.}
Our approach applies sequence-to-sequence learning through a recurrent neural network (RNN) encoder-decoder model to automatically learn assert statements within test methods. This model is inspired by Chen \etal \cite{Chen:2019}, which attempts to predict a single line of code that has a predetermined place holder within the method. The goal of this deep learning strategy is to learn a conditional distribution of a variable length sequence conditioned on a completely separate variable length sequence $P(y_1, y_2,\dots, y_m | x_1, x_2,\dots, x_n)$. Where $n$ and $m$ may differ. %In many cases, $n$ and $m$ are different lengths. %The model handles this by creating a fixed length vector to represent both the input sequence and the output sequence.
During training, the model's encoder is fed the tokenized input sequence of the test method plus the context of the focal method as a single stream of tokens ($x_1, \dots, x_n$). The assert statement, which is our target output sequence, has been removed and replaced with a specialized token. The decoder attempts to accurately predict the assert ($y_1, \dots, y_m$) by minimizing the error between the decoder's generated assert and the oracle assert statement. This is accomplished by using the negative log likelihood of the target tokens using stochastic gradient descent \cite{Kiefer1952StochasticEO}. An overview of an RNN encoder-decoder can be seen in \figref{fig:approach}.

\begin{figure}
	\centering
	\includegraphics[width=\linewidth]{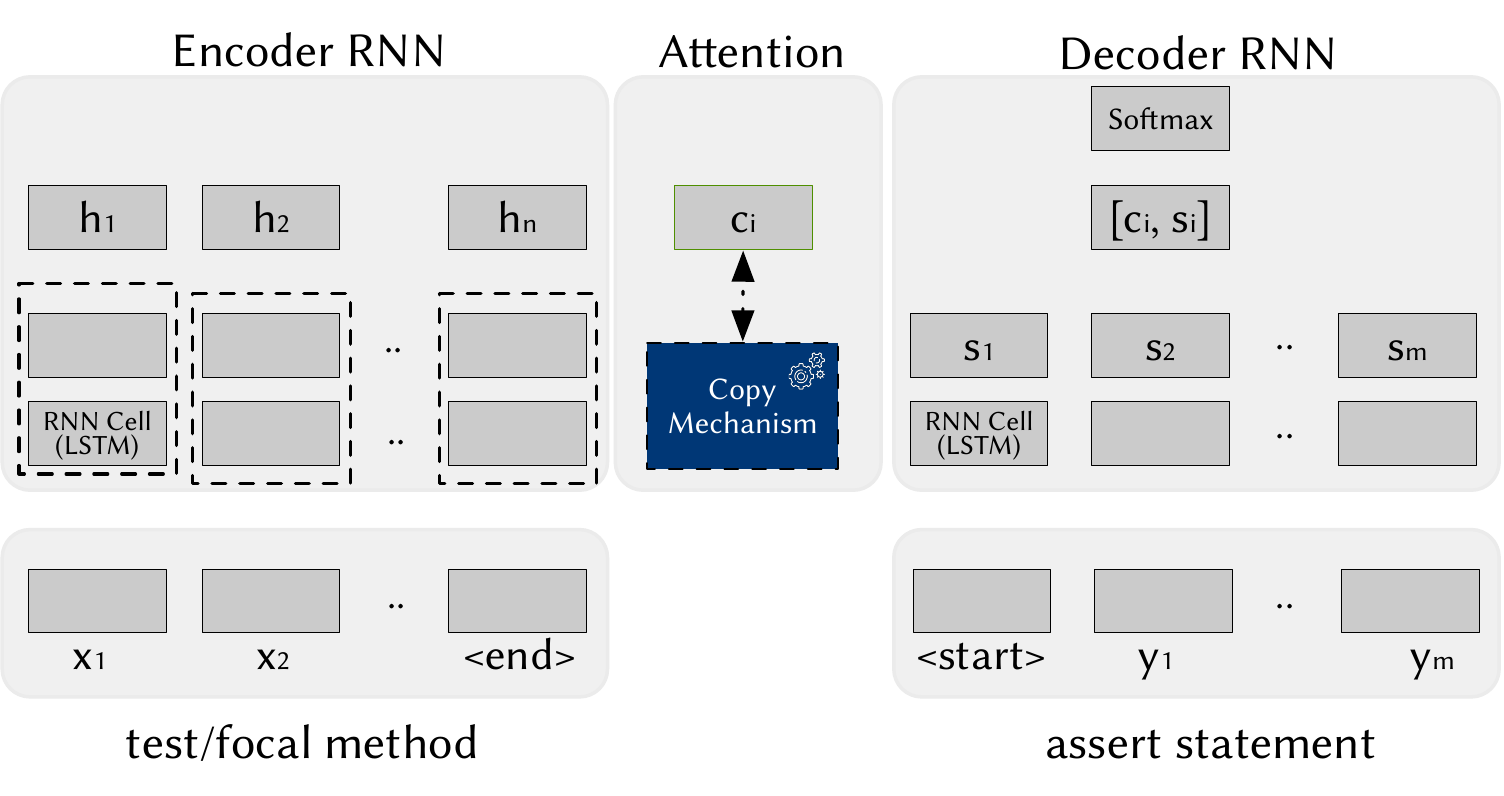}
	\vspace{-0.8cm}
	\caption{The RNN bidirectional encoder decoder model}
	\vspace{-0.4cm}
	\label{fig:encoder_decoder}
\end{figure}

%\begin{figure}
%	\centering
%	\includegraphics[width=\linewidth]{fig/encoder_decoder.png}
%	\caption{The RNN bidirectional encoder decoder model \MICHELE{I would personally drop this image to make space for the approach one. Encoder-Decoder models are now pretty %well known also in SE and the written explanation should be enough.}}
%	\label{fig:encoder_decoder}
%\end{figure}

\vspace{-0.2cm}
\subsection{Encoder}
\label{sec:approach:encoder}
The encoder is a single layer bi-directional RNN, which is comprised of two distinct LSTM RNNs. This bidirectionality allows the encoder to consider both the tokens that come before and the tokens that come after as context for the token of interest. The encoder takes a variable length sequence of source code tokens $X = (x_1, x_2,\dots, x_n)$ as input. From these tokens, the encoder produces a sequence of hidden states $(h_1, h_2,\dots, h_n)$ generated from LSTM RNN cells. These cells perform a series of operations to propagate relevant information, including previous hidden states, to the next cell. Due to the bidirectionality of the encoder, there exists a sequence of hidden states when considering the token sequence from left to right $\overrightarrow{h_i} = f(x_{i}, h_{i-1})$ and right to left $\overleftarrow{h_i} = f(x_{i_0}, h_{{i_0}+1})$. For our model, each hidden state can be formally described as the non-linear activation of the current sequence token and the previously synthesized hidden state. Once the hidden state for each directional pass is found, they are concatenated to derive the finalized hidden state $h_i$. The sequence of resulting hidden states is propagated through the model as the context vector. The encoder also applies the regularization technique of dropout at a rate of $0.2$.     

\subsection{Attention Mechanism}
\label{sec:approach:attention}
The context vector $\mathcal{C}$, commonly referred to as an attention mechanism, is computed as a weighted average of the hidden states from the encoder $\mathcal{C} = \sum_{i=1}^{n} \alpha_i h_i$. In this equation $\alpha$ represents a vector of weights used to denote the influence of different parts of the input sequence. In this manner, the model can pay greater \textit{attention} to particular tokens of the input sequence when attempting to predict the output token $y_i$. The weights which influence the attention mechanism are trained over time to help identify the patterns of contribution from different input tokens.  

\subsection{Decoder and Copy Mechanism}
\label{sec:approach:decoder}
The decoder is a double layer LSTM RNN that learns to take a fixed length context vector and translates it into a variable length sequence of output tokens. Given a previous hidden state $h_{\hat{i}-1}$, the previous predicted token $y_{\hat{i}-1}$ and the context vector $\mathcal{C}$ the decoder generates a new hidden state that can be used to predict the next output token. As done for the encoder, we apply regularization to the decoder by using dropout at a rate of $0.20$, and the Adam optimizer for learning with a starting learning rate of 0.0001:
\begin{equation*}
	h_{\hat{i}} = f(h_{\hat{i}-1}, y_{\hat{i}-1}, \mathcal{C})
\end{equation*}
The decoder generates a new hidden state each time it predicts the next token in the sequence until a special stop token is reached. The hidden states are generated using the equation above. However, these hidden states are also used by the copy mechanism to help predict the appropriate output token. In particular, the copy mechanism works to calculate two separate probabilities. The first, is the probability that the next predicted token in the output sequence should be taken from the vocabulary. The second, is the probability that the next predicted token in the output sequence should be \textit{copied} from the input sequence. With the ability to consider copying tokens from the input sequence to the output sequence, we can artificially extend the vocabulary of the encoder-decorder RNN model for the raw dataset. Each sequence inferred from the model now has the ability to consider any predetermined vocabulary token, in addition to any token from the input sequence. The downside is that the copy mechanism is a trainable extension of the model, that learns which input tokens should be copied over. The benefit is that the model can better deal with rare tokens that would otherwise not appear in the vocabulary.  

\begin{figure}
	\centering
	\includegraphics[width=1.02\linewidth]{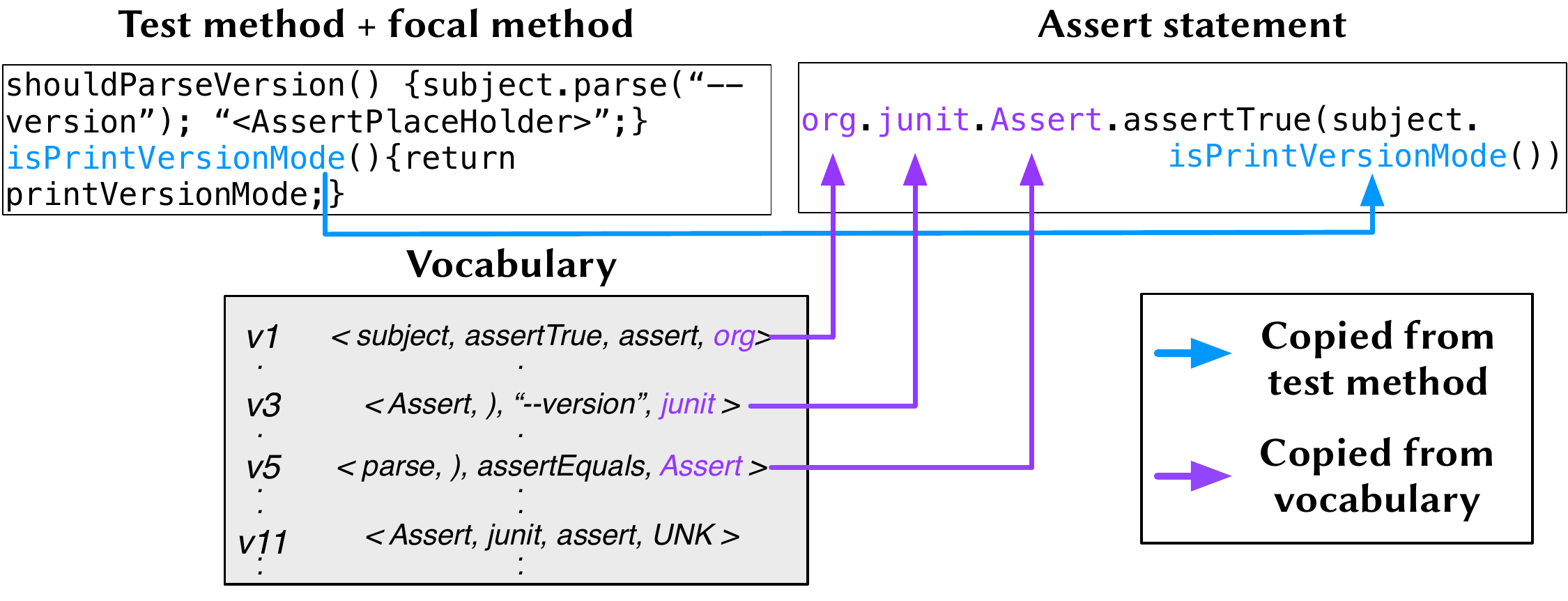}
	\vspace{-0.5cm}	
	\caption{Copy Mechanism Example}
	\vspace{-0.6cm}
	\label{fig:copy_mech}
\end{figure}

To further demonstrate how the copy mechanism works, consider \figref{fig:copy_mech}. In this example, we see the vocabulary tokens predicted for each time step $v_t$. For the purpose of this example we do not consider the translation of the separator tokens. However, in a real scenario, our model would predict these tokens in identical fashion. The first predicted token for the output sequence is the \texttt{org} token, which is copied from the vocabulary. This process is repeated for the \texttt{junit} token, the \texttt{Assert} token and would continue for all further tokens but the last one (\ie \texttt{isPrintVersionMode}). The last token is not found anywhere in the vocabulary. At time step $v_{11}$ the model recommends the \texttt{UNK} token to indicate that the highest probability token does not exist within the defined vocabulary. In this case, the copy mechanism is used to determine which input token has the highest probability to be the predicted token. In the case shown in the example, the context appended from the focal method contains this token and it is copied to the output sequence. Without the copy mechanism, there would have been no way to resolve this example and it would have been discarded. It is important to note that the copy mechanism is trained along with the network and requires no effort on the end users.    

Also, note that the copy mechanism is only applied to the raw dataset of source code tokens. The reasoning is that after abstracting the source code, all tokens are available within the vocabulary. Therefore, the probability that a predicted token would be outside of the vocabulary, and within the input sequence, is 0\%, rendering the copy mechanism useless. Rather, our abstraction process serves as a pseudo copy mechanism of typified IDs. Consider the previous example in \figref{fig:abstraction} where \texttt{isPrintVersionMode} was not found anywhere in the vocabulary. In our abstraction process, this token is abstracted into METHOD\_1 and since METHOD\_1 is a term contained within our vocabulary the model has no problem in predicting this token in the output assert statement. Then, when we map the abstracted assert statement back to raw source code, we replace METHOD\_1 with the token \texttt{isPrintVersionMode} without relying on the copy mechanism.
%\MICHELE{Good explanation of the intuition behind the copy mechanism, I also like the figure and the example. However, I think we need to spend a bit more words to explain why the abstracted corpus does not need the copy mechanism. Our abstraction process together with the code concretization process, basically represents a form of source-code-specific copy mechanism. Because, instead of having \texttt{unk} tokens that are copied, we have typified IDs that are copied. If possible, I would add exactly the same example in Figure 4 but with the abstracted code. In that case, the token isPrintVersionMode is abstracted as "METHOD\_1" and therefore will be copied in the assert, later using the mapping we can replace the ID.}
	
\section{Experimental Design}
\label{sec:experimental_design}
% Experiment Design

The \emph{goal} of our study is to determine if \atlas can generate meaningful, syntactically and semantically correct assert statements for a given test method and focal method context. Additionally, it is important for our approach to be lightweight in order to require as little overhead as possible if, for example, it is combined with approaches for the automatic generation of test cases. %Therefore, we empirically assess these three components of whether NMT is a viable approach for generating meaningful assert statements.

%\KEVIN{I think we should calrify here that this is the test set of the data, and that the training set is separate and give the size. This will make things more clear.}
The \emph{context} of our study is represented by a dataset of 158,096 TAPs for the abstracted dataset and 188,154 TAPs for the raw dataset. These datasets are further broken down into 126,477 TAPs for training, 15,809 TAPs for validation, and 15,810 TAPs for testing in the abstract dataset. Likewise, we had 150,523 TAPs for training, 18,816 TAPs for validation and 18,815 TAPs for testing in our raw dataset. The differences in number of examples between the two datasets is due exclusively to the removal of duplicates. Since the abstracted model reuses typified IDs, there is a greater chance for the duplication of TAPs within the abstracted dataset. %All TAPs were mined from GitHub as discussed in \secref{sec:approach:github}.
Our evaluation aims at answering the research questions described in the following paragraphs. \smallskip

\textbf{RQ$_1$: Is A{\footnotesize TLAS} able to generate assert statements resembling those manually written be developers?} We assess whether \atlas is a viable solution for generating semantically and syntactically correct assert statements. Therefore, we perform experiments on real-world test methods and determine if our model can predict the appropriate assert statement. We use both datasets (\ie raw source code and abstracted code) to train the encoder-decoder recurrent neural network model. During training, we use our validation set to determine the optimal parameterization of the NMT model (complete list of used parameters available in \cite{replication}). We then evaluate the trained model on the test set, which contains examples previously unseen in both the training set and the validation set.

We begin training our model on TAPs, feeding the model the test method and associated focal method context. We train our model until the evaluation on the validation set shows that the models parameterization has reached a (near-)optimal state (\ie the model is no longer improving the calculated loss for data points outside the training set). This is a common practice to prevent the effects of overfitting to the training data. Our training phase results in two separate models, one for predicting raw source code assert statements and the other for predicting abstracted asserts. Remember that when working with raw source code, we also implement the copy mechanism, which is not used for abstracted code. In total, the abstract model trained for 34 hours while the raw model trained for 38 hours. The difference in training time can be attributed to the use and training of the copy mechanism in conjunction with the lack of abstraction. 

Once the model is trained, inference is performed using beam search \cite{Raychev:2014:CCS:2594291.2594321}. The main intuition behind beam search decoding is that rather than predicting at each time step the token with the best probability, the decoding process keeps track of $k$ hypotheses (with $k$ being the beam size). Thus, for a given input (\ie test method + focal method), the model will produce $k$ examples of assert statements. We experiment with beam sizes going from $k=1$ to $k=50$ at steps of 5.

Given the assert statements predicted by our approach, we consider a prediction as correct if it is identical to the one manually written by developers for the test method provided as input. We refer to these asserts as ``perfect predictions''. When experimenting with different beam sizes, we check whether a perfect prediction exists within the $k$ generated solutions. We report the raw counts and percentages associated with the number of perfect predictions. % that our model can generate.

Note that, while the perfect predictions certainly represent cases of success for our approach, this does not imply that the ``imperfect predictions'' all represent failure cases (\ie the generated asserts are not meaningful). Indeed, for the same test/focal method, different assert statements could represent a valid solution. Therefore, we sample 100 ``imperfect predictions'' and manually analyze them to understand whether, while different from the original assert statements written by the developer, they still represent a meaningful prediction for the given test method. In particular, we split the ``imperfect predictions'' into four sets based on their BLEU-4 score \cite{Papineni:2002} value. The BLEU score is a well-known metric for assessing the quality of text automatically translated from one language to another \cite{Papineni:2002}. In our case, the two ``languages'' are represented by i) the test method and the focal method, and ii) the generated assert statement. We use the BLEU-4 variant, meaning that the BLEU score is computed by considering the 4-grams in the generated text, as previously done in other software-related tasks in the literature \cite{Gu:2016:DAL:2950290.2950334,Jiang:ASE'17}. The BLEU score ranges between 0\% and 100\%, with 100\% indicating, in our case, that the generated assert is identical to the reference one (\ie the one manually written by developers). We use the BLEU score ranges 0-24, 25-49, 50-74 and 75-99 to split the imperfect predictions. Then, we randomly selected 25 instances from each set and the first author manually evaluated them to determine if the generated assert statement is meaningful in the context of the related test/focal methods. To avoid subjectiveness issues, the 100 instances were also randomly assigned to four other authors (25 each) who acted as second evaluator for each instance. Conflicts (\ie cases in which one of the two evaluators classified the assert statement as meaningful while the other did not) arose in a single case that was solved through an open discussion. We report the number of meaningful assert statements we found in the manually analyzed sample as empirical evidence that imperfect assert statements could still be useful in certain cases. Note that, while there might be authors' bias in assessing the meaningfulness of the ``imperfect'' assert statements (\ie authors may tend to be too positive in evaluating the meaningfulness of the asserts), we make our evaluation publicly available in the replication package \cite{replication}, to allow the reader to analyze the performed classification. \smallskip

\textbf{RQ$_2$: Which types of assert statements is A{\footnotesize TLAS} capable of generating?} After obtaining the perfect predictions from RQ$_1$, we analyze the types of assert statements our model can generate. In particular, we analyze the taxonomy of assert statements generated by our approach in the context of perfect predictions to determine the types of assert statements the model is able to correctly predict. We then report the raw counts and the percentages for each type of assert statement the model can perfectly predict. 

Note for this evaluation, we only report results for $k=1$, since we found that already with this beam size the model was able to generate all types of assert statements in the used dataset.  \smallskip

\textbf{RQ$_3$: Does the abstraction process aid in the prediction of meaningful assert statements?} %In this RQ, we want to determine if the abstraction process allows to improve the prediction performance of \atlas. 
Remember that while our abstraction model generates abstracted asserts, we can map them back to the raw source code at no cost to the developer. Thus, we can check whether the generated assert is a perfect prediction, as we do for the raw source code. Besides comparing the performance of the two models, we also analyze whether they produce complementary results by computing the following overlap metrics:

$$
pp_{R \cap A} = {|pp_{R} \cap pp_{A}| \over |pp_{R} \cup pp_{A}|}\hspace{0.2cm}
pp_{R \setminus A} = {|pp_{R} \setminus pp_{A}| \over |pp_{R} \cup pp_{A}|}\hspace{0.2cm}
pp_{A \setminus R} = {|pp_{A} \setminus pp_{R}| \over |pp_{R} \cup pp_{A}|}
$$
%\smallskip

The formulas above use the following metrics: $pp_{R}$ ($pp_{A}$) represents the set of perfect predictions generated using the raw (abstracted) source code dataset;
$pp_{R \cap A}$ measures the overlap between the set of perfect predictions generated by using the two datasets;
$pp_{R \setminus A}$ measures the perfect predictions generated on the raw source code but not when using abstraction (\emph{vice versa} for $pp_{A \setminus R}$).

%\begin{itemize}
%\item  $pp_{R}$ ($pp_{A}$) represents the set of perfect predictions generated using the raw (abstracted) source code dataset;
%\item $pp_{R \cap A}$ measures the overlap between the set of perfect predictions generated by using the two datasets;
%\item $pp_{R \setminus A}$ measures the perfect predictions generated on the raw source code but not when using abstraction (\emph{vice versa} for $pp_{A \setminus R}$).
%\end{itemize}
\smallskip

\textbf{RQ$_4$: What is the effect of using the copy mechanism in our model?} With the addition of the copy mechanism we can perfectly predict assert statements which contain tokens only found in the input sequence and not in the vocabulary. Here we want to quantify the effect of the copy mechanism and see how many assert predictions we would be capable of producing without its usage. Therefore, we analyze the perfect prediction set from the raw source code model. If the perfect prediction contains a token not found in the vocabulary, then we know that its generation was possible due to the usage of the copy mechanism. We report the raw counts and percentages of the number of assert statements that were resolved thanks to the use of the copy mechanism. \smallskip

\textbf{RQ$_5$: Does A{\footnotesize TLAS} outperform a baseline, frequency-based approach?} As output of RQ$_2$ we defined a taxonomy of assert statements that our approach is able to correctly generate in the perfect predictions. We noted that there are eight types of assert statements that the model is capable of generating. However, the model also predicts the variables and method calls contained within the assert statement. Therefore, we want to determine if the most frequently used assert statements found within our dataset (\eg \emph{assert(true)}), could be used to create a frequency-based approach that outperforms our learning-based approach. 

We analyze the duplicated assert statements generated in our perfect prediction set. This means that the model generated the same (correct) assert statement for different test methods provided as input. It is important to highlight here that while the same assert statement can be used in different test methods, this does not imply that we have duplications in our datasets. Indeed, while the same assert statement can be used in different TAPs, the test/focal methods are guaranteed to be unique.

The developed frequency-based approach takes the most commonly found assert statements and applies them as a solution for a given test method. In particular, we take the top $k$ most frequent assert statements and test if any of them represents a viable solution for each test method in the test set. In this evaluation we set the same $k$ for both approaches (\ie the frequency-based and the learning-based ones). For example, assuming $k=5$, this means that for the frequency-based approach we use the five most frequent assert statements as predictions, while for the learning-based approach we set beam size to five. We report the raw counts of the frequency-based approach as compared to our NMT-based approach. We compare the two approaches at $k=1,5,10$.\smallskip

\textbf{RQ$_6$: What is the inference time of the model?} Our last research question speaks to the applicability and ease of use of our model. When developing test cases within a software project, it is unreasonable to expect the developer to spend a considerable amount of time to set up and run an inference of the model. Therefore, we performed a timing analysis to assess the time needed to generate assert statements for a variety of beam sizes. %For the training time, we recorded how long it took to train each model given the specified parameters of the model and the number of data points in our dataset. For analyzing the inference time we used $k=1$ to $k=50$ with an increment of 5 to test the trade off between the timing of the model's inference and the increased prediction results of the model.
In particular, we used $k=1$ to $k=50$ with an increment of 5 to test the trade off between the timing of the model's inference and the increased prediction results of the model.

% Perfect Prediction Table

\begin{table}[]
	\vspace{0.7cm}
	\caption{Prediction Classification}
	\label{tab:perfect_pred}
	\resizebox{\linewidth}{!}{
		\begin{tabular}{|c|c|c|c|c|}
			\hline
			\multirow{2}{*}{\textbf{Beam Size}} & \multicolumn{2}{c|}{\textbf{Raw Model}}                                                                                                                         & \multicolumn{2}{c|}{\textbf{Abstract Model}}                                                                                                                    \\ \cline{2-5} 
			& \textbf{\begin{tabular}[c]{@{}c@{}}Peferct Prediction\\ Percentage\end{tabular}} & \textbf{\begin{tabular}[c]{@{}c@{}}Perfect Prediction\\ Counts\end{tabular}} & \textbf{\begin{tabular}[c]{@{}c@{}}Perfect Prediction\\ Percentage\end{tabular}} & \textbf{\begin{tabular}[c]{@{}c@{}}Perfect Prediction\\ Counts\end{tabular}} \\ \hline
			1                                   & 17.66\%                                                                          & 3323                                                                         & 31.42\%                                                                          & 4968                                                                         \\ \hline
			5                                   & 23.33\%                                                                          & 4390                                                                         & 49.69\%                                                                          & 7857                                                                         \\ \hline
			10                                  & 24.73\%                                                                          & 4654                                                                         & 55.73\%                                                                          & 8812                                                                         \\ \hline
			15                                  & 25.53\%                                                                          & 4805                                                                         & 58.76\%                                                                          & 9291                                                                         \\ \hline
			20                                  & 25.88\%                                                                          & 4871                                                                         & 60.43\%                                                                          & 9554                                                                         \\ \hline
			25                                  & 26.19\%                                                                          & 4929                                                                         & 61.75\%                                                                          & 9764                                                                         \\ \hline
			30                                  & 26.43\%                                                                          & 4973                                                                         & 62.73\%                                                                          & 9918                                                                         \\ \hline
			35                                  & 26.63\%                                                                          & 5012                                                                         & 63.68\%                                                                          & 10068                                                                        \\ \hline
			40                                  & 26.81\%                                                                          & 5045                                                                         & 64.38\%                                                                          & 10179                                                                        \\ \hline
			45                                  & 26.91\%                                                                          & 5064                                                                         & 64.81\%                                                                          & 10247                                                                        \\ \hline
			50                                  & 27.01\%                                                                          & 5083                                                                         & 65.31\%                                                                          & 10327                                                                        \\ \hline
		\end{tabular}
	}
\vspace{-0.5cm}
\end{table}

We record the results in number of seconds and map the increased performance against the increased time. We do not consider the time it may take for a developer to look at all resulting predictions of assert statements for different beam sizes. Note that we do not consider the training time since this is a one time cost that does not affect the usability of the approach. %This is a limitation of the approach, which we address in our threats to validity section. 
%\MICHELE{I think we all agree that we could sacrifice this RQ in case we need more space for qualitative examples. We can summarize the timing with a short sentence and without graphs.}

\section{Experimental Results}
\label{sec:results}
% Results
%We discuss the results aimed at answering our formulated research questions. 

\vspace{1mm}\textbf{RQ$_1$ \& RQ$_3$ \& RQ$_4$: Ability to generate meaningful assert statements, comparison between raw and abstracted dataset, and usefulness of copy mechanism.} \tabref{tab:perfect_pred} shows the perfect prediction rate for both the ``raw model'' and the ``abstract model'' for beam sizes 1 and 5-50 at increments of 5 (RQ$_1$ \& RQ$_3$). As expected, the perfect prediction rate increases using larger beam sizes, with a plateau reached at beam size 20 (\ie only minor increases in performance are observed for larger beam sizes).

When using beam size equal to 1 for the model trained/tested with the raw dataset, our approach generates 17.66\% perfect predictions, resulting in over 3.3k correctly generated assert statements. The average BLEU score for the asserts predicted when only considering the top recommendation (\ie beam size = 1) is 61.85. We also found that the copy mechanism helps the raw model in generating perfect predictions (RQ$_4$). Indeed, we determined how many perfect predictions require the use of the copy mechanism, finding that, when using beam size equals 1, our approach is able to perfectly predict 3323 examples by using the copy mechanism, and 2439 when only relying on the vocabulary. This means that the copy mechanism is responsible for resolving 884 perfect predictions, which constitutes 4.69\% of the perfect prediction rate.

For the abstract model, the percentage of perfect predictions goes up to 31.42\% ($\sim$5k correctly generated asserts). Here the average BLEU score is 68.01. Note that we mention the BLEU scores for completeness, but do not perform a full evaluation using this metric. 

\begin{figure}[t]
	\centering
	\vspace{0.4cm}
	\includegraphics[width=0.79\columnwidth]{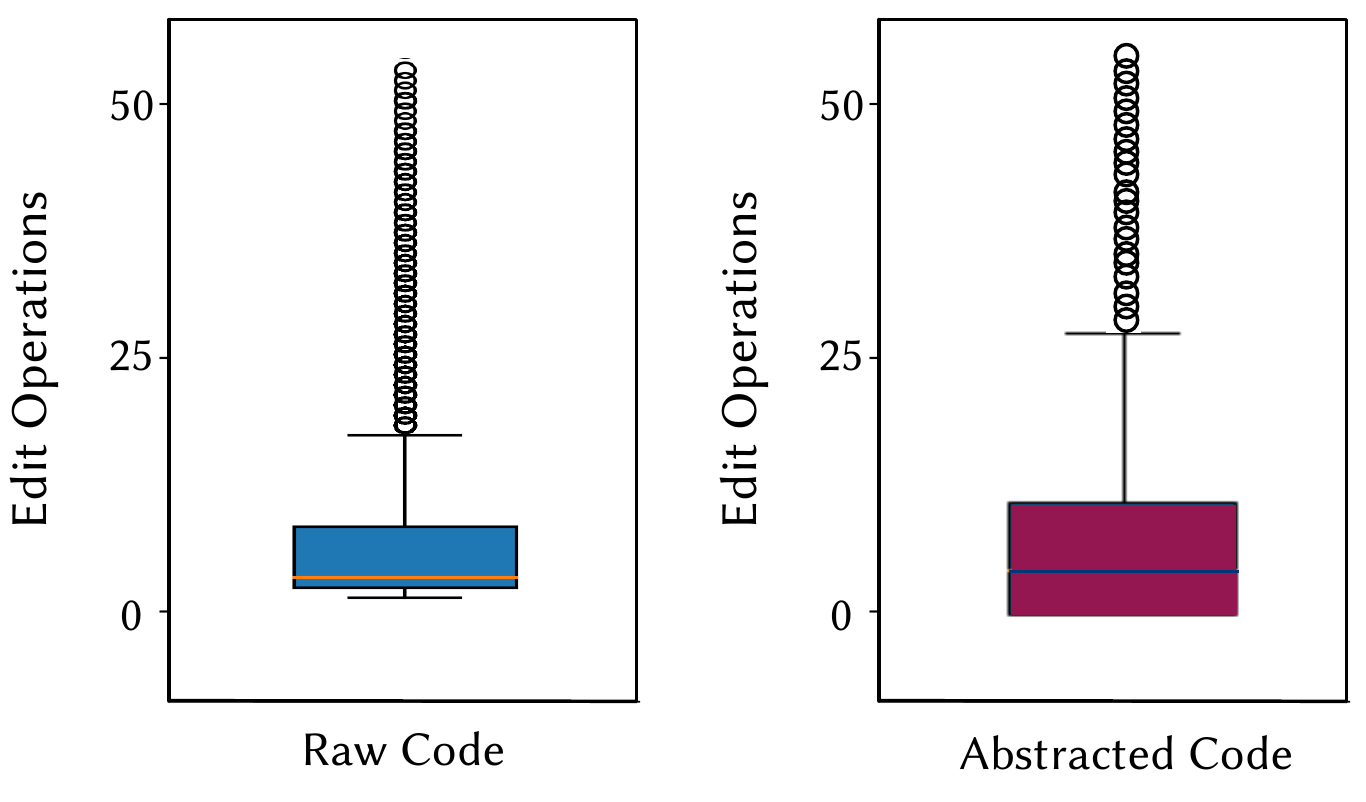}
	\vspace{-0.4cm}
	\caption{Edit Distance between Imperfect Predictions and Ground Truth (Truncated tail of higher edit distances)}	
\label{fig:editdist}
\vspace{-0.6cm}
\end{figure}

Of particular note is the substantial bump in perfect predictions that we obtain as result of increasing the beam size to five for both models, especially for the abstracted one. For the latter, in 49.69\% of test/focal methods provided as input, one of the generated asserts is identical to the one manually written by the developer (for a total of 7.9k perfectly predicted asserts). This indicates the potential of our approach in an automatic ``code completion'' scenario, in which the developer could easily pick one of the five recommended asserts. 

\begin{figure*}[t]
	\centering
	%\vspace{-0.2cm}
	\includegraphics[width=\linewidth]{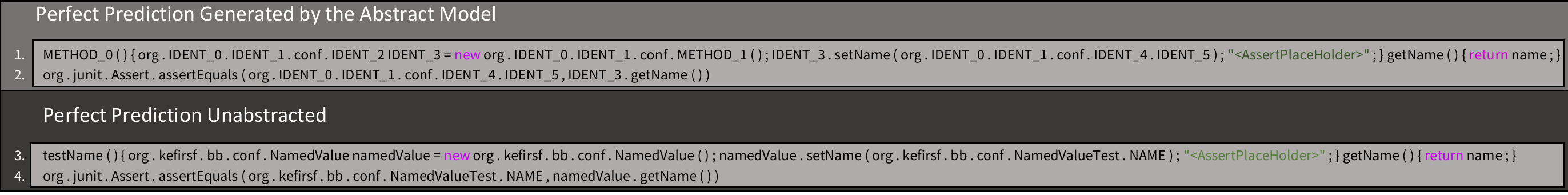}
	\vspace{-0.6cm}
	\caption{Example of a Perfect Prediction Made by the Abstract Model, and its unabstructed mapping to actual code}	
\label{fig:qualexample}
\vspace{-0.3cm}
\end{figure*}

As mentioned in \secref{sec:experimental_design}, we also analyze the complementarity of the perfect predictions generated by the raw and the abstracted models when using beam size equals 1. In terms of overlap (\ie $pp_{R \cap A}$) we found that only 117 examples were captured and perfectly predicted by both models. Also, 3206 (39.2\%) of the perfect predictions are only generated by the raw model ($pp_{R \setminus A}$), while 4851 (59.3\%) can only be obtained by using the abstracted model ($pp_{A \setminus R}$). This shows a surprising complementarity of the two models, that are able to learn different patterns in the data. The combination of these two different representations, for example by using a model with two encoders (one per representation) and a single decoder, may be an interesting direction for future work.

However, the overall higher performance in both BLEU score and perfect prediction rate of the abstract model shows that it is better able to \textit{translate} a meaningful assert statement from the test and focal method context. This is likely due to the different way in which the two models see the out-of-vocabulary tokens. In the raw dataset they are all represented as unknown tokens, while in the abstracted dataset they are represented with the associated type (\eg METHOD, VAR, INT, \etc). When dealing with out-of-vocabulary tokens, this helps the abstracted model in learning patterns such as: a "METHOD" token is likely to follow a "VAR" and a "." token. For the raw model, this only results in observing ``UNK . UNK'', hindering its ability to learn meaningful patterns in these cases. Due to the better performance ensured by the abstract model, the subsequent analyses are performed using this model. %\MICHELE{Technically, we analyze again the raw model with copy later on.}

Here, we discuss an interesting example generated by our abstracted model. Figure \ref{fig:qualexample} presents the abstracted contextual method in line 1 and the abstracted  assert statement in line 2. Lines 3 and 4 are the result of the unabstraction process, where we map back the abstracted tokens into raw source code. In this example, the test method is creating a new object \textit{namedValue} and setting the \textit{NAME} attribute of this object. The test method contains the method call \textit{getName}, which our heuristic identifies as the focal method and is appended to the end of the test method. The model then generates an assert statement that compares the \textit{NAME} attribute of the new object with the results from the \textit{getName} method call. Indeed, the model learns the relationship between the test and focal method in order to generate a meaningful assert statement which appropriate tests the method's logic. This assert statement is a nontrivial example of the model determining what type of assert statement to generate, as well as the appropriate parameters the assert statement should contain.

%Since the model learns a probability over the distribution of tokens, the difference in the two model's performance can be expected. Abstraction artificially reduces the vocabulary much more than the raw dataset. \MICHELE{IMPORTANT: This is actually wrong. The abstraction process does not reduce the vocabulary much more than the raw dataset. Indeed, it's the opposite! The raw vocabulary is 1k tokens, while the abstract vocabulary is 1k tokens (idioms) + a bunch of IDs ($\sim$100 for METHOD1,2,3... and all the types.). The main difference is that, out-of-vocabulary tokens are treated differently. The raw dataset sees ALL of them as "unk" tokens. While the abstracted dataset see them with the associated type (METHOD, VAR, INT, etc.), this provides a huge help in learning patterns such as: after "VAR" and "." there is probably a "METHOD". Instead, the raw dataset sees them as: unk . unk } Although the copy mechanism does aid in the ability to use rare tokens during the inference phase, the model is responsible for tuning and learning the weights associated with the copy mechanism. The abstracted dataset has a much lower vocabulary, even with the inclusion of idioms, thereby reducing the strain on the model to tune weights associated with a copy mechanism.

Concerning the manual inspection of the 100 ``imperfect predictions'', we found that 10\% of them can represent a valuable assert statement for the provided contextual method, despite them being different from the asserts manually written by developers. Note that, while a 10\% might look like a ``negative'' result, it is important to understand that these results are in addition to the already perfectly predicted assert statements. %Also, this evaluation was strict in what was considered valuable and it is possible that developers could glean logic or important details from seemingly incorrect assert statements. 
The full evaluation of the 100 cases is available in our replication package \cite{replication}. Here we discuss a representative example, for a provided test/focal method, where our approach generated the assert statement \texttt{assertSame\-(input\-,result)} while the assert manually written by the developer was \texttt{assertEquals\-("Blablabla"\-,result)}. The \texttt{input} object is present in the test method, and contains indeed the value "Blablabla". Basically, our model generated an ``easier to maintain'' assert, since changes to the value assigned to the input objects do not require changes to the assert statement. Concerning the difference between \texttt{assertSame} and \texttt{assertEquals}, the former compares two objects using the \texttt{==} operator, while the latter uses the \texttt{equals} method that, if not overridden, does also perform the comparison using the \texttt{==} operator.

Although we have shown that our model can produce meaningful results outside of the perfect predictions, we wanted to understand ``how far'' are the imperfect predictions from the manually written assert statements. Therefore, for each imperfect prediction generated by the abstract model with $k=1$, we computed the number of tokens one needs to change, add, or delete to convert it into the manually written assert. We found that by only changing one token it is possible to convert 23.62\% of the imperfect assert statements (\ie 3660 instances) into perfect predictions. Also, the median of 3 indicates that in half of the cases, changing only three tokens would be sufficient. Note that the average number of tokens in the generated assert statements is 17.1. \figref{fig:editdist} shows the distribution of edit changes needed for the imperfect predictions to become perfect predictions.

\vspace{-0.2cm}
\begin{center} 
\fbox{
\begin{minipage}[t]{0.97\linewidth}
{\bf Summary for RQ$_1$, RQ$_3$, \& RQ$_4$.} 
\atlas's abstracted model is able to perfectly predict assert statements (according to a developer written ground truth) 31.42\% and 49.69\% of the time for top-1 and top-5 predictions respectively. Conversely, the model operating on raw source code with the copy mechanism achieved a perfect prediction rate of 17.66\% and 23.33\% for top-1 and top-5 predictions respectively. This indicates the abstracted model performs better overall. However, we also found that models had a relatively high degree of orthogonality, with 39.2\% of all perfect predictions generated by the raw model, and 59.3\% generated by the abstracted model, illustrating that the copy-mechanism allowed for the prediction of a unique set of assert statements.%We are able to determine that NMT is a viable tool when attempting to generate meaningful assert statements for developer written test methods. Additionally, we performed a comparison demonstrating the differences between the use of abstraction versus the use of raw tokens in conjunction with a copy mechanism. We determined that the abstracted data set performed better overall, achieving a perfect prediction rate for 31.42\% of our test cases.
\end{minipage}
}
\end{center}
 
% Assert Types

\begin{table}[]
	\centering
	\footnotesize
	\caption{Types of Predicted Assert Statements}
	\vspace{-0.2cm}
	\label{tab:assert_types}
	%\resizebox{\linewidth}{!}{%
		\begin{tabular}{|c|c|c|}
			\hline
			\textbf{Assert Type} & \textbf{Count} & \textbf{Total in Dataset} \\ \hline
			assertEquals         & 2518           & 7923                      \\ \hline
			assertTrue           & 973            & 2817                      \\ \hline
			assertNotNull        & 606            & 1175                      \\ \hline
			assertThat           & 250            & 1449                      \\ \hline
			assertNull           & 238            & 802                       \\ \hline
			assertFalse          & 232            & 1017                      \\ \hline
			assertArrayEquals    & 102            & 311                       \\ \hline
			assertSame           & 49             & 314                       \\ \hline
		\end{tabular}%
	\vspace{-0.3cm}
\end{table}

% Frequency based table

\begin{table}[]
	\centering
	\footnotesize
	\caption{Learning Based vs. Frequency Based}
	\vspace{-0.3cm}
	\label{tab:freq}
	%\resizebox{\linewidth}{!}{%
		\begin{tabular}{|c|c|c|}
			\hline
			\multirow{2}{*}{\textbf{Beam Size}} & \multicolumn{2}{c|}{\textbf{Number of Perfect Predictions}} \\ \cline{2-3} 
			& \textbf{Frequency Model} & \textbf{Abstract Model} \\ \hline
			1 & 455 &  4968 \\ \hline
			5 & 970 &  7857 \\ \hline
			10 & 1299 & 8812 \\ \hline
		\end{tabular}%
	%}
\vspace{-0.4cm}
\end{table}  

\vspace{1mm}\textbf{RQ$_2$: Which types of assert statements is ATLAS capable of generating?} 
%After assessing that capability of ATLAS of generating meaningful assert statements, 
We also analyzed the types of JUnit4 assert statements that were perfectly predicted. Here we use the 4,968 perfect predictions generated using the abstract model and beam size equals one. \tabref{tab:assert_types} shows that our approach was able to correctly predict eight different types of assert statements, with \texttt{assertEquals} and \texttt{assertTrue} being the most commonly predicted type of assert statement. Note that some JUnit4 assert types (\eg \texttt{assertNotSame} and \texttt{assertFail}) are not generated by our model because they were not present in our dataset.

As it can be seen in \tabref{tab:assert_types}, the distribution of assert statements we are able to predict is similar to the one of the assert statements in the entire dataset. This result mitigates the possible threat that our approach is only successful in generating a specific type of assert statement. Indeed, as shown in \tabref{tab:assert_types}, the lack of a uniform distribution of data points in the dataset from which \atlas is learning seems to be the main reason for the skewed distribution of correctly predicted asserts. The main exception to this trend is represented by the \texttt{assertThat} statements. We hypothesize that \texttt{assertThat} statements are more difficult to predict due to the nature of the assert itself. These types of asserts compare a value with a matcher statement, which can be quite complex, since matcher statements can be negated, combined, or customized. Despite the complexities of \textit{assertThat} statements, the model is still able to perfectly predict 17.2\% of the ones seen in the testing set. 

\vspace{-0.2cm}
\begin{center} 
\fbox{
\begin{minipage}[t]{0.97\linewidth}
{\bf Summary for RQ$_2$.} 
We found that \texttt{assertEquals} was the most common type of assert generated, matching the distribution we were learning from. Our analysis showed that \atlas is capable of learning every type of assert statement found in developer written test cases. 
\end{minipage}
}
\end{center}

%\input{tab/frequency_based}  

%%%%%

%\input{tab/copy_mechanism}

\vspace{1mm}\textbf{RQ$_5$: Does ATLAS outperform a baseline, frequency-based approach?} In this research question we explore whether our abstract model outperforms a baseline, frequency-based approach (see \secref{sec:experimental_design}). \tabref{tab:freq} shows the results of this comparison. We note that our DL-based approach outperforms the frequency based approach at each experimented beam size. The difference in terms of performance is substantial, resulting in 6.8 (beam size=10) to 10.9 (beam size=1) times more perfect predictions generated by our approach. For example, when only considering the top candidate assert statement for both techniques, \atlas correctly predicts 4968 assert statements, as compared to the 455 of the frequency-based model. The achieved results indicate that \atlas is in fact learning relationships based on hierarchical features and not being overwhelmed by repetitive assert statements. We also want to note that \atlas encompassed a majority of the assert statements found by the frequency based baseline. Therefore, we do not combine a frequency based approach with \atlas and believe our implementation to be superior on its own.

\vspace{-0.2cm}
\begin{center} 
\fbox{
\begin{minipage}[t]{0.97\linewidth}
{\bf Summary for RQ$_5$.} 
\atlas is able to significantly outperform a frequency-based baseline prediction technique by a factor of 10.
\end{minipage}
}
\end{center}

%\vspace{1mm}\textbf{RQ$_6$: Are assert statements predicted useful and meaningful in a real-world software development scenario?} \GAB{I skipped this RQ for now.} For this RQ, we attempt to understand how useful our approach is in a real world development scenario. We extract test methods and generate assert statements for the test methods we extract. We record the comments and actions taken by developers pertaining to the pull requests we create with the newly synthesized assert statements. 

\begin{figure} 
	\centering
	%\vspace{-0.2cm}
	\includegraphics[width=0.9\columnwidth]{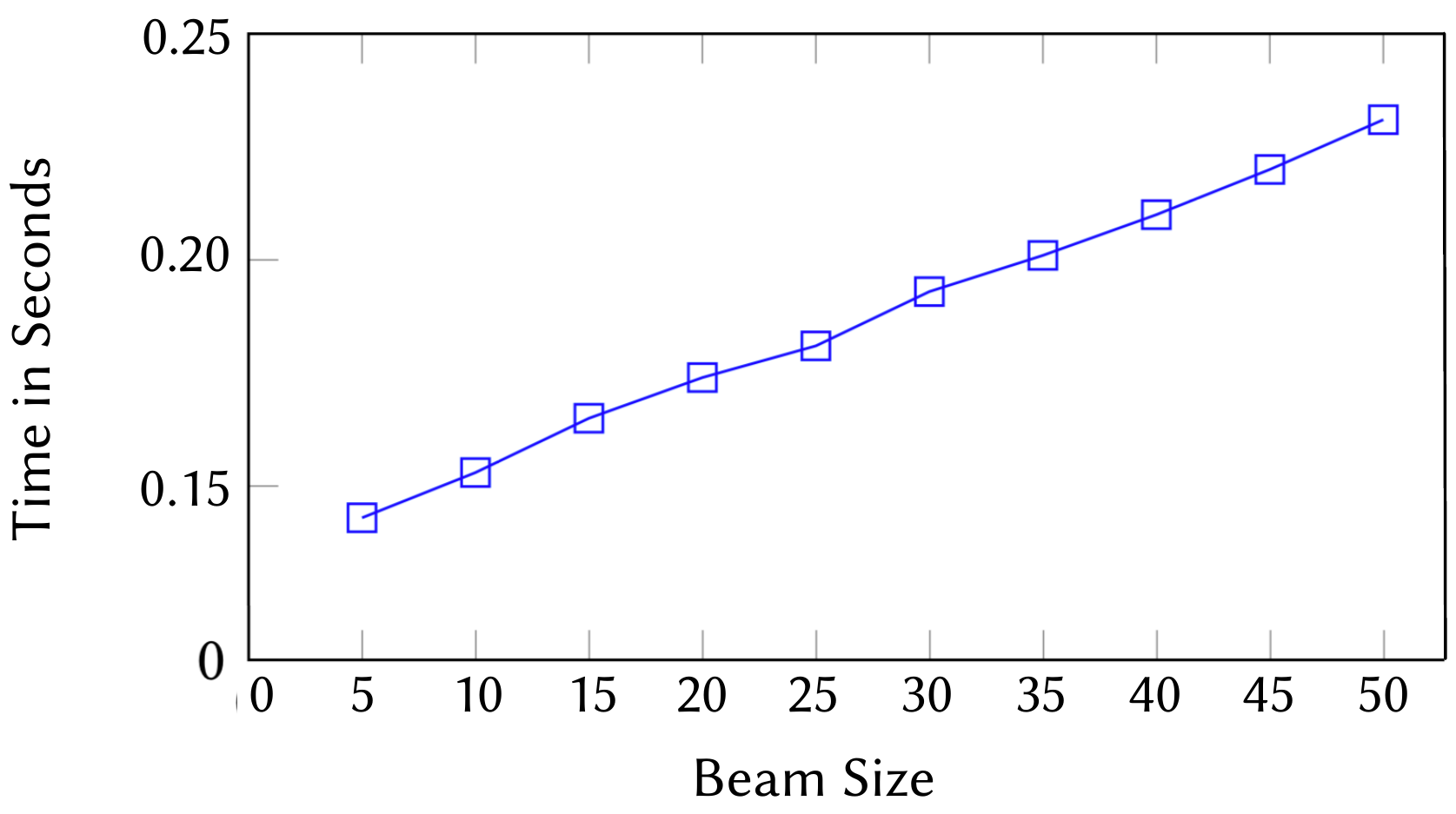}
	\vspace{-0.3cm}
	\caption{Seconds per Assert Generation}	
\label{fig:timings}
\vspace{-0.6cm}
\end{figure}

\vspace{1mm}\textbf{RQ$_6$: What is the inference time of the model?} Our last research question assesses the time required by our approach to generate meaningful assert statements. Given the previously discussed performance of the experimented models, we computed the generation time on the abstract model. It is important to note the reported time does not include the code abstraction nor the mapping of the prediction back to source code, although these operations are typically more efficient than inference. \figref{fig:timings} shows the increase in time based on the number of solutions the model generates. These timings concern the generation of assert statements for 15,810 test/focal methods provided as input. The reported time is in seconds per provided input, and includes the generation of all assert statements in the given beam size. For example, assuming a beam size of 5, it would take around 0.14 seconds to generate all 5 predictions for a particular test/focal method. These results were computed using a single consumer-grade NVIDIA RTX GPU.

\begin{center} 
\fbox{
\begin{minipage}[t]{\linewidth}
{\bf Summary for RQ$_6$.} 
We found that with a beam size of 5, we can generate all predictions in 0.14 seconds per test method $+$ focal method pairing. When increasing beam size, we find that the time needed to generate assert statements appears to scale linearly.
\end{minipage}
}
\end{center}

\section{Threats to Validity}
\label{sec:threats}
% Threats to Validity

\textbf{Construct validity} threats concern the relationship between theory and observation, and are mainly related to sources of imprecision in our analyses. We automatically mined and parsed the TAPs used in our study. In this process, the main source of noise is represented by the heuristic we used to identify the focal method for a given assert statement. As said, in a real usage scenario, this information could be provided by the developer who wrote the test or by the automatic test case generation tool. Despite the presence of introduced noise, our approach was still able to generate meaningful asserts, confirming the robustness of our NMT model.

\textbf{Internal validity} threats concern factors internal to our study that could influence our results.  The performance of our approach depends on the hyperparameter configuration, that we report in our online replication package \cite{replication}. However, given how computationally-expensive the hyperparameter search was, we did not investigate the impact of the copy mechanism across all configurations. 

\textbf{External validity} threats concern the generalizability of our findings. We did not compare ATLAS with state-of-the-art test case generation techniques using the heuristics described in \secref{sec:motivation} to define appropriate asserts. This comparison would require a manual evaluation of the correctness of the asserts generated by ATLAS and by the competitive techniques for automatically generated tests, likely on software of which we have little knowledge (assuming Open-Source projects). Furthermore, we would have to make the subject systems executable (e.g., as required by EvoSuite) which is known to be difficult. Hence, this would not allow us to scale such an experiment to the size of our current evaluation.  Indeed, our main goal was to empirically investigate the feasibility of a learning-based approach for assert statement generation. Comparing/combining the two families of approaches is part of our future work. Finally, we only focused on Java programs and the JUnit framework. However, ATLAS's learning process is language-independent and its NMT infrastructure can easily be ported to other programming languages.

\section{Conclusion and Future Work}
\label{sec:conclusion}
% Conclusion

In this work, we mined over 9k GitHub projects to identify test methods with their related assert statements. We then used a heuristic to identify the focal method associated with an assert statement and generated a dataset of Test-Assert Pairs (TAPs), composed by i) an input sequence of the test and focal method, and ii) a target sequence reporting the appropriate assert statement for the input sequence. We used these TAPs to create a NMT-based approach, called ATLAS, capable of learning semantically and syntactically correct asserts given the context of the test and focal methods. 

We found that the model was capable of successfully predicting over 31\% of the assert statements developers wrote by only using the top-ranked prediction. When looking at the top-5 predictions the percentage of correctly generated assert statements grew up to $\sim$50\%. We also showed that among the ``imperfect predictions'', meaning the scenario in which ATLAS generates assert statements different from the ones manually written by developers, there are assert statements that either represent a plausible alternative the the one written by the developer or can be converted into the latter by just modifying a few tokens ($\leq 3$ in 50\% of cases). Finally, some of the limitations of our approach, as well as the extensive empirical study we conducted, provide us with a number of lessons learned that can drive future research in the field:

\noindent{\emph{\textbf{Raw code vs. Abstracted code}}}: Our results show that through the abstraction mechanism, applications of NMT on code can ensure better performance, as already observed in previous studies \cite{DBLP:journals/corr/abs-1901-09102, DBLP:journals/corr/abs-1812-10772, Tufano:2018:EIL:3238147.3240732}. More interestingly, we found that the two code representations are quite complementary, and allow our approach to generate different sets of ``perfect predictions''. This points to the possibility of combining the two representations into a single model that could benefit from an architecture having two encoders (one per representation) and a single decoder.

\noindent{\emph{\textbf{On the possibility of generating multiple assert statements}}: We investigated this research direction while working on ATLAS. The main problem we faced was the automatic identification of the part of test method body that it is relevant for  a given assert. Indeed, while with a single assert we can assume that the whole method body is relevant, this is not the case when dealing with multiple asserts. Here, a possible heuristic could be to consider the statements preceding the assert statement $a_1$, but coming after the previous assert $a_0$, as relevant for $a_1$. However, it is unlikely that the statements coming before $a_0$ are totally irrelevant for $a_1$. Another possibility we considered was to apply backward slicing on each assert statement but, unfortunately, this resulted in scalability issues and in (well-known) problems related to the automatic compilation of open source projects \cite{Tufano:JSEP'16}. Approaching this problem is a compelling direction for future work.}

\noindent{\emph{\textbf{Integrating a learning-based approach in tools for automatic test case generation}}: As discussed earlier, we foresee two possible usages for ATLAS. First, it can be used as a code completion mechanism when manually writing unit tests in the IDE. Second, it could be combined with existing tools for automatic test case generation \cite{Fraser:ESEC/FSE'11,radoop,agitar}. We currently lack empirical evidence to substantiate any claim on the effectiveness of ATLAS in improving automatically generated tests, which would be an important first step prior to integration. Additionally, before combining ATLAS with existing tools, it is necessary to deeply understand the cases in which the two families of approaches (\ie the ones integrated in the test case generation tool and the learning-based one) succeed and fail. In this way, learning-based approaches could be used only when needed (\ie when the standard approach implemented in the test case generation tools is likely to fail), thus increasing the effectiveness of the generated tests. Studying and comparing the strengths and weakness of the two families of techniques is part of our future research agenda.}

The automatic generation of meaningful assert statements is a compelling problem within the software engineering community. We showed that a learning-based approach could help aid in this problem, and opened a complementary research direction to the one already adopted in automatic test case generation tools \cite{Fraser:ESEC/FSE'11,radoop,agitar}.

\begin{acks}
This work is supported in part by the NSF CCF-1927679 and CCF-1815186 grants. Bavota acknowledges the financial support of the Swiss National Science Foundation for the project CCQR (SNF Project No. 175513).
\end{acks}

%\clearpage

\bibliographystyle{ACM-Reference-Format}
\bibliography{main}

\end{document}